\documentclass[12pt]{article}
\usepackage{latexsym}
\usepackage{amsfonts}
\makeatletter

\def\tr{\mathop{\hbox{\rm tr}}\nolimits}

\def\be{\begin{equation}}
\def\ee{\end{equation}}
\def\ba{\begin{array}}
\def\ea{\end{array}}
\def\bea{\begin{eqnarray}}
\def\eea{\end{eqnarray}}
\def\dd{\partial}

\def\one#1{#1^{\raise5pt\hbox{$\scriptstyle\!\!\!\!1$}}\,{}}
\def\two#1{#1^{\raise5pt\hbox{$\scriptstyle\!\!\!\!2$}}\,{}}

\makeatletter
\def\binrel@#1{\begingroup
  \setboxz@h{\thinmuskip0mu
    \medmuskip\m@ne mu\thickmuskip\@ne mu
    \setbox\tw@\hbox{$#1\m@th$}\kern-\wd\tw@
    ${}#1{}\m@th$}%
  \edef\@tempa{\endgroup\let\noexpand\binrel@@
    \ifdim\wdz@<\z@ \mathbin
    \else\ifdim\wdz@>\z@ \mathrel
    \else \relax\fi\fi}%
  \@tempa
}
\let\binrel@@\relax
\def\overset#1#2{\binrel@{#2}%
  \binrel@@{\mathop{\kern\z@#2}\limits^{#1}}}
\def\underset#1#2{\binrel@{#2}%
  \binrel@@{\mathop{\kern\z@#2}\limits_{#1}}}
\makeatother
\newfont{\bbd}{msbm10 scaled\magstep1}

\begin{document}
%\hfill NTZ 28/1999
\begin{center}
{\Large{Zero-curvature condition in Calogero model}}
\end{center}
\begin{center}

\end{center}
\begin{center}
D.~Karakhanyan,${}^{a,b}$\footnote{e-mail: karakhan@yerphi.am},
 Sh. Khachatryan${}^{a}$\  \\

\bigskip

{\em ${}^a$Yerevan Physics Institute\\ 2 Alikhanyan Brothers St.,
Yerevan 0036, Armenia}

\bigskip
{\em ${}^b$Yerevan State University\\
1 Alex Manoogian Str., Yerevan, 0025, Armenia\\}
\end{center}

\begin{abstract}
We consider the mutual commutativity of Dunkl operators of the
rational Calogero model as zero-curvature condition and calculate
the non-local operator, related to these flat connections. This
operator has physical meaning of particular scattering matrix of
Calogero model and maps the eigenfunctions of Dunkl operator to the
wave function of $N$ free particles (plane waves).
\end{abstract}

%%%%%%%%%%%%%%%%%%%%%%%%%%%%%%%%%%%%%%%%%%%%%%%%%%%%%%%%%%%%%%%%%%%%%%%%%%%%%%

%{\small \tableofcontents}
\renewcommand{\refname}{References.}

\renewcommand{\theequation}{\thesection.\arabic{equation}}
%\setcounter{equation}{0}

%\section{Introduction}
%\setcounter{equation}{0}
\section{Introduction}
Quantum Calogero-Moser-Sutherland model describes the set of $N$
identical particles on a circle interacting pairwise with inverse
square potential. Calogero-Moser-Sutherland model attracted some
attention due to conformal character of interaction potential, it
also used to test the ideas of fractional statistics \cite{Ha},
\cite{Hal}. The new aspects of their algebraic structure and quantum
integrability was later clarified by \cite{p} and \cite{He}. The
Calogero Model \cite{cal} (see for review \cite{p}) is a rare
example of integrable many-body problem. The studying of the
Hamiltonian $H_c$ was started by Calogero \cite{cal}, who computed
the spectrum, eigenfunctions and scattering states in the confined
and free cases. The Perelomov \cite{Pe} observed the complete
quantum integrability of the model, he stated that exist $N$
commuting, algebraically independent operators and $H_c$ is one
among them. The complete integrability of the classical Hamiltonian
of Calogero-Moser-Sutherland was proved by Moser \cite{Mo}.

Unfortunately, in a series of integrable theories, Calogero model
stands alone. A powerful Inverse Scattering Method \cite{kbi} is
applicable to it with considerable restrictions. The key object in
ISM is $R$-matrix depending on spectral parameter. The R-matrix can
associate to this model, but it is dynamic: its matrix elements are
not c-numbers and depend on the coordinates. Moreover, the
dependence on the spectral parameter can be obtained only for
elliptical extension \cite{skl}. It makes us look for other ways to
describe this model.

The Lax method being applied to the integrable model allows to trace
its integrability to zero-curvature condition of $L-A$ pair. Namely
the equations of motion of the model express the consistency
condition of some (usually more wide) linear (free) system.

The Calogero model in external harmonic field
 \be \label{Calogero}
H_C=\sum_{i=1}^N \left(\frac{p_i^2}2+\frac{\omega^2q_i^2}2\right)
+\sum_{i<j}\frac{g^2}{(q_i-q_j)^2}
 \ee
can be obtained by $SU(N)$ reduction from the matrix model
 \be\label{P2+Q2}
H=\frac12\tr{P^2}+\frac{\omega^2}{2}\tr{Q^2}
 \ee
where $Q$ and $P$ are hermitean matrices:
 \be\label{pq}
\{P_{ij},Q_{i',j'}\}=\delta_{ij'}\delta_{ji'}, \quad
\{P_a,Q_b\}=\delta_{ab},
 \ee
which is equivalent to the homogeneous $(N^2-1)$-dimensional
oscillator.

Consider the Calogero matrix Hamiltonian without oscillator term
$\omega=0 $. Then (\ref{P2+Q2}) tell us $H_0=\frac12tr(P^2)$. The
canonical relations (\ref{P2+Q2}) are invariant under (canonical)
similarity transformation:
$$
P\to P'=U^{-1}PU,\qquad Q\to Q'=U^{-1}QU,
$$
with numerical unitary matrix $U=\exp{i\varepsilon}$. The Noether
current corresponding to this transformation is $tr(P\delta Q)$ for
infinitesimal $\varepsilon$ one has $\delta Q=i[Q;\varepsilon]$ and
$tr(P\delta Q)=tr(i\varepsilon[P;Q])$. So one deduces, that the
Noether charges for this transformation are:
 \be\label{jij}
J_{ij}=i[P;Q]_{ij}=const.
 \ee
Now, using that symmetry one can turn the matrix $Q$ to diagonal
form:
 \be\label{q}
Q=diag(q_1,q_2,\ldots,q_N).
 \ee
Then one can define matrix $P$ from (\ref{jij}):
 \be\label{p}
J_{ij}=i\sum_j( P_{ik}q_k\delta_{kj}-q_i\delta_{ik}P_{kj}) =iP_{ij}
(q_j-q_i),\qquad\Rightarrow\qquad P_{ij}=\frac{iJ_{ij}}{q_i-q_j}.
 \ee
Thus the result of reduction of matrix model under consideration is
entirely determined by the numerical matrix $J$ with zeros at
diagonal. The simplest case of zero (rank) matrix $J=0$ corresponds
to diagonal matrix $P$ (diagonal elements can not be derived from
(\ref{p}) due to vanishing the diagonal elements of $J$) and to the
case of $N$ free particles.

The next in complexity case is rank two matrix:
 \be\label{j2}
J_{ij}=\delta_{ij}-w_iw_j,\qquad w_i=1,\quad i=1,\ldots N,
 \ee
the rank one matrix is excluded due to vanishing diagonal elements.
Let $\psi_i$ are components of column $\psi$ eigenvector
corresponding to eigenvalue $\lambda$:
$$
\psi_i-\sum_{j=1}^N\psi_j=\lambda\psi_i,\qquad\Rightarrow\qquad(1-
\lambda)\psi_i=(1-\lambda)\psi_j,
$$
one sees, that eigenvalue $\lambda_1=1$ has multiplicity $N-1$,
while eigenvalue $\lambda_2=1-N$ has multiplicity one, i.e. the
matrix (\ref{pij}) indeed has rank two.

So finally, matrix $P$ is:
 \be\label{pij}
P_{ij}=p_i\delta_{ij}+i\frac{1-\delta_{ij}}{x_i-x_j}=\left\{\ba{cc}
p_i,\qquad\;i=j\\\frac i{q_i-q_j},\quad i\neq j\ea\right.
 \ee

Substituting (\ref{p}) for this case into (\ref{P2+Q2}) one will
come to Hamiltonian of Calogero model:
$$
H=trP^2=\sum_{i,j=1}^NP_{ij}P_{ji}=\sum_{i=1}^NP_{ii}^2+2\sum_{i<j}
P_{ij}P_{ji}=\sum_{i=1}^Np_i^2+2\sum_{i<j}\frac1{(x_i-x_j)^2}.
$$
Then the $N-1$ integrals of motion are given by $I_j=trP^j$,
$j\neq2$. In particular,
 \be\label{p3}
tr P^3=\sum_{i=1}^Np_i^3+3\sum_{i,j=1}^N\frac{1-\delta_{ij}}{x_i-x_
j}p_i\frac1{x_i-x_j}-i\sum_{i,j,k=1}^N\frac{(1-\delta_{ij})(1-
\delta_{jk})(1-\delta_{ki})}{(x_i-x_ j)(x_j-x_k)(x_k-x_k)}
 \ee

\section{Dunkl operator}
%\subsection{}
However, the reduction is not the only way to prove the
integrability of Calogero model. In this article we will try to
trace the origin of integrability in approach, associated with the
permutations of the particles \cite{dunkl}. The motivation for the
introduction of Dunkl operators served as their relationship with
the Calogero model and quantum many body systems of
Calogero-Moser-Sutherland type. Introduced in this way Dunkl
operators are commuting differential-difference operators, related
to a finite reflection group on a Euclidean space. First the class
rational operators was introduced by C.F. Dunkl in a series of
papers \cite{dun}. He also introduced the framework for a theory of
special functions and integral transforms in several variables
related with permutation groups. Then the various other classes of
Dunkl operators were invented such that trigonometric Dunkl
operators of Heckman, Opdam and the Cherednik operators \cite{hec}.

Dunkl operator, relevant for rational Calogero model ia $A_{N-1}$
type (is related to $GL(N)$ group). The key object of this approach
is permutation operator:
 \be\label{pik}
{\mathcal{P}}_{ik}\cdot f(\ldots,x_i,\ldots,x_k,\ldots)=
f(\ldots,x_k,\ldots ,x_i,\ldots),
 \ee
where $f$ is an arbitrary smooth function of $N$ variables. It has a
number of obvious properties, that are easy to follow from the
definition (\ref{pik}):
 \be\label{piks}
{\mathcal{P}}_{ik}={\mathcal{P}}_{ki},\qquad
{\mathcal{P}}_{ik}{\mathcal{P}}_{ik}=1,
 \ee
symmetry
 \be\label{pikc}
{\mathcal{P}}_{ij}{\mathcal{P}}_{kl}={\mathcal{P}}_{kl}{\mathcal{P}}
_{ij},\qquad i\neq j\neq k\neq l,
 \ee
commutativity and
 \be\label{pikf}
{\mathcal{P}}_{ik}{\mathcal{P}}_{kl}={\mathcal{P}}_{il}{\mathcal{P}}
_{ik}={\mathcal{P}}_{kl}{\mathcal{P}}_{il},\qquad i\neq k\neq l,
 \ee
fusion. The Dunkl operator, playing the central role in our
consideration is defined as follows:
 \be\label{m1}
\nabla_k=\dd_k-c\sum_{i\neq
k}\frac1{x_i-x_k}{{\mathcal{P}}_{ik}}=\dd_k-c\sum_{i=1}^N\frac{1-\delta_{ik}}{x
_i-x_k}{{\mathcal{P}}_{ik}}\equiv\dd_k-{\mathcal{A}}_k.
 \ee
It is closely related with Calogero Hamiltonian $H_c$:
$$
Res(\sum_{i=1}^N\nabla_i^2)=-2H_c,\qquad H_c=-\frac12\Delta+\sum_{
i<j}^N\frac{c(c-1)}{(x_i-x_j)^2},
$$
where the symbol $Res(A)$ means the restriction of operator $A$ to
the space of invariants of permutation group $S_N$. In other words,
under that sign the permutation operator ${\mathcal{P}}_{ik}$ at
utmost right position can be replaced by unity.

Namely the sum of squared Dunkl operators differs from the
Calogero-Moser Hamiltonian only in term, linear by coupling constant
$c$:
 \be\label{n2}
\sum_{i=1}^N\nabla_i^2=\sum_{i=1}^N\left(\dd_i^2+c\sum_{j=1}^N\frac
{1-\delta_{ij}}{(x_i-x_j)^2}{\mathcal{P}}_{ij}+c^2\sum_{j=1}^N\frac
{1-\delta_{ij}}{(x_i-x_j)^2}\right).
 \ee
This relation suggests that the totally symmetric and totally
antisymmetric combination of the eigenfunctions of Dunkl operators
(on which permutation in second term takes values $\pm1$) can serve
as the wave functions of Calogero model.

This property emphasizes the connection between hamiltonian of
Calogero model and Dunkl operators, but from the point of view of
integrability the next property is even more important:
\section{Zero-curvature condition}

Following the invention of the Dunkl operator, it was realized that
its components commute:
 \be\label{dd}
[\nabla_j,\nabla_k]=0.
 \ee
The zero curvature condition (\ref{dd}) expresses the integrability
of Calogero model: in the functional space, where the non-local
connections (like ${\mathcal{A}}_k$) are allowed, it is equivalent
to free (non-interacting) model.

Thus, in this context, the non-local gauge field appears differently
from the usual non-local field theory, where the non-locality is
allowed in the microscopic region of space in order to avoid
divergences. Thus, on a macroscopic scale causality is not violated
\cite{ef}. In contrast, in the present context model under
consideration, although the nature of the nonlocal interactions is
integrable.

In order to formulate this observation more correct, we will solve
zero curvature condition above introducing non-local operator $U$:
 \be\label{u}
\nabla_k=U^{-1}\dd_kU,\qquad {\mathcal{A}}_k=-U^{-1}(\dd_kU),
 \ee
here bracket means that derivative $\dd_k$ acts only on variables
$x_k$ contained in operator $U$, but not on test function $\psi(x)$
in defining relation:
 \be\label{df}
{\mathcal{A}}_k\cdot\psi=-[U^{-1}(\dd_kU)]\cdot\psi=[(\dd_kU^{-1})
U]\cdot\psi.
 \ee
In order to represent the formal solution to (\ref{df}) as a
path-ordered exponential, we consider an arbitrary smooth curve
$(BB')$:
 \be\label{curve}
x_k=x_k(t), \qquad B=(x_1(0),\ldots,x_N(0)),\;\;B'=(x_1(t'),\ldots,
x_N(t')).
 \ee
Multiplying (\ref{df}) by $\dot x_k$ and summing us  by $k$ one
obtains:
 \be\label{up}
[\dot U^{-1}U]\cdot\psi=\frac12\sum_{i,k=1}^N(1-\delta_{ik})\frac
{\dot x_i-\dot x_k}{x_i-x_k}{\mathcal{P}}_{ik}\cdot\psi\equiv
{\bf{A}}\cdot\psi.
 \ee
The all possible pairwise transpositions of the arguments of test
function $\psi$ stand in r.h.s. of (\ref{up}). Then one can formally
write:
 \be\label{au}
\dot U^{-1}\cdot\psi={\bf{A}}(t)U^{-1}\cdot\psi,
 \ee
 \be\label{au}
U^{-1}(t)=1+\int_0^t{\bf{A}}(t_1)dt_1+\int_0^t\int_0^{t_1}{\bf{A}}
(t_2)dt_2{\bf{A}}(t_1)dt_1+\ldots,
 \ee
indeed, taking derivative in r.h.s. one factors out ${\bf{A}}$ and
reproduces whole series, i.e. $U^{-1}$. This series is properly
defined when ${\bf{A}}$ is given by function of $t$, but when
${\bf{A}}$'s given by operators, as in (\ref{au}), their product
should be ordered.  Before define that, two notations are in order.

First is related to singularities of Calogero Hamiltonian at
$x_i=x_k$. We can avoid them restricting ourself by considering of
some simplex, say $x_1<x_2<\ldots<x_N$ instead of whole space and
demand that curve $x_k(t)$ belongs to that simplex, but careful
analysis \cite{tsu} shows that correct choice of boundary conditions
for wave function when approaching to singular point leads to
consistent quantization scheme. The good illustration is the $N=2$
case, after excluding the center of mass, it reduces to the study of
the 1-dimensional Schr\"{o}dinger operator:
 \be
H_x=-\frac{\hbar^2}{2m}\frac{d^2}{dx^2}+\frac m2\omega^2x^2+\frac g2
x^{-2}.
 \ee
The spectrum of $H_y$ is unbounded from below if $g<-\frac{\hbar^2}
{4m}$ and Calogero assumed in his work that $g>-\frac{\hbar^2}{4m}$.
For "admissible" wave functions he imposed the constraint that
associated probability current vanish at the point where any two
particles collide. The selection of admissible wave functions is
equivalent to choosing a domain on which the Hamiltonian is
self-adjoint. Then, in \cite{tsu1} shown that there exits a family
of different possibilities parameterized by a $2\times2$ unitary
matrix if $g<\frac{3\hbar^2}{4m}$. In the corresponding
quantizations of the $N=2$ Calogero model the probability current
does not in general vanish at the coincidence of the coordinates of
the particles.

Since this phenomenon refers to the interaction of any pairs of
particles, one may expect it to occur also in the $N$ particle
Calogero model. Moreover, the case $N=3$ considered in \cite{tsu1}
in details.

The second notation consists of operator-valued nature of eq.
(\ref{au}). The operator ${\bf{A}}$ can be represented as a function
of $2N$ variables $x_k$ and $\dd_k$:
 \be\label{syp}
({\bf{A}}\cdot f)(x)=a(x,\dd)f(x)=\frac1{(2\pi h)^{N/2}}\int d^Npe^
{\frac{ix\cdot p}h}a(x,\frac ihp)f(p).
 \ee
Here $a(x,\frac ihp)$ is symbol of operator ${\bf{A}}$ and it has to
be represented in ordered form, which can be unambiguously restored
from symbol. For example the normal ordering (all $x$'s stand from
the left and derivatives stand from the right) can be chosen. This
dependence on differential operators in ${\bf{A}}$ comes of course
from permutation ${\mathcal{P}}_{jk}$. It can be expressed in many
equivalent ways:
 \be\label{perm}
\!\!\!\!{\mathcal{P}}_{jk}\!=(-1)^{x_k\dd_k}e^{x_j\dd_k} e^{-x_k
\dd_j} e^{x_j\dd _k}\!=e^{x_k\dd_j}e^{-x_j\dd_k} e^{x_k
\dd_j}(-1)^{x_k\dd_k}\!=(-1)^{\frac12(x_k-x_j)(\dd_k-\dd_j)}.
 \ee
However we need the normally ordered expression for permutation
operator:
 \be\label{taylor}
(-1)^{v\dd_v}\cdot f(v)=f(-v)=\sum_{n=0}^\infty\frac{(-2v)^n}{n!}
\dd_v^n\cdot f(v),
 \ee
in order to construct its symbol.

In Appendix C we prove the equivalence of different realization for
$\mathcal{P}_{ik}$ (\ref{perm}).

So, the symbol of permutation operator can be written as:
 \be\label{sym}
Symbol[{\mathcal{P}}_{jk}]=a(x,\frac ihp)=\sum_{n=0}^\infty\frac{(
x_k-x_j)^n}{n!}(-\frac ih(p_k-p_j))^n.
 \ee
Together with formulae (\ref{syp}) and (\ref{au}) it defines the
formal series for operator $U^{-1}$.
\section{Eigenproblem}
In order to clarify the physical meaning of operator $U$ it makes
sense to define eigenproblem of Dunkl operators:
 \be\label{nk}
\nabla_k\psi(x_1,x_2,\ldots,x_N)=ip_k\psi(x_1,x_2,\ldots,x_N),
 \ee
here imaginary unit is extracted in order to eigenvalue $p_k$ will
real. Due to zero curvature condition (\ref{dd}) equations above
compatible each to other. Another, non-trivial integrability
condition to this set of equations provides permutation operator.
Due to relations:
 \be\label{ic}
{\mathcal{P}}_{jk}\nabla_k=\nabla_j{\mathcal{P}}_{jk},\qquad
[{\mathcal{P}}_{jk};\nabla_l]=0,\quad j\neq l,\;k\neq l,
 \ee
proven in Appendix B, one deduces, that if $\psi_{p_1,p_2,\ldots,p_N
}(x_1,x_2,\ldots,x_N)$ is solution to the set (\ref{nk}), then the
function
$$
{\mathcal{P}}_{jk}\psi_{p_1,p_2,\ldots,p_N}(x_1,x_2,
\ldots,x_N),
$$
which differs from
$$
\psi_{p_1,p_2,\ldots,p_N
}(x_1,x_2,\ldots,x_N)
$$
by permutation of arguments $x_k$ and $x_j$
will be solution to the same equation with permutation of parameters
$p_k$ and $p_j$. Taking into account, that the set od $N$ linear
first order differential equations has a unique solution, this
condition can be formulated as follows:
 \be\label{sc}
\psi_{\ldots p_j,\ldots,p_k,\ldots}(\ldots x_j,\ldots,x_k,\ldots)=
\psi_{\ldots p_k,\ldots,p_j,\ldots}(\ldots x_k,\ldots,x_j,\ldots).
 \ee
In other words, this restriction reduces the arbitrariness in the
choice of function of $2N$ variables to one totally symmetric
function:
$$
\psi(f_1,\ldots,f_N),\qquad \psi(\ldots,f_j,\ldots,f_k,\ldots)=
\psi(\ldots,f_k,\ldots,f_j,\ldots),\qquad f_k\equiv f(x_k,p_k),
$$
where $f(x,p)$ is an arbitrary function of two variables.

Let us multiply eqs. (\ref{nk}) by $U$:
 \be\label{un}
U\nabla_k\psi=\dd_kU\psi=ip_kU\psi,
 \ee
from which one deduces that if solution to eqs. (\ref{nk}) is given
by function $\psi$, then the function $\psi_0=U\psi$ satisfies to
the set of equations:
$$
\dd_k\psi_0=ip_k\psi_0,
$$
which has unique solution:
 \be\label{p0}
\psi_0=const\cdot e^{i(x_1p_1+x_2p_2+\ldots+x_Np_N)}.
 \ee
In the spirit of remark about connection between totally symmetric
and totally antisymmetric combinations of $\psi$'s and Calogero wave
functions, one sees, that the operator $U^{-1}$ maps symmetrized,
according to principle of identity of particles, combination of
$\psi_0$'s to Calogero wave function with $g=c(c+1)$ and $g=c(c-1)$.

In this regard, the physical meaning of operator consists of
$U^{-1}$ is just scattering matrix of the rational Calogero model.

So, another way to determine operator $U^{-1}$ consists of solving
the set of equations (\ref{nk}) and restoring $U^{-1}$ by function
$\psi$.

\section{Conclusion and outlook}
The operator $U$, introduced as a solution to the condition of zero
curvature associated with the non-local gauge transformation that
reflects the wave function of the Calogero model in the wave
function of $N$ non-interacting particles may be of interest from
the point of view of non-commutative geometry, because it provides
the realization of micro-causality condition within a particular
model of interacting particles.

Also of interest is a generalization of this consideration to the
case of the anionic particle statistics, which was discussed by
A.P.Polychronakos \cite{p}.

\section*{Acknowledgements} This work is supported in part by Armenian
grants: SCS 13-1C132, ANSEF grant mathph 3122, D.K. is supported
also by grants: SCS 13RF-018 (2013) and Volkswagen Foundation I/84
496. Authors are grateful to Tigran Hakobyan for valuable
discussions and to Armen Allahverdyan for careful reading of
manuscript.

\section{Appendix A}
In this Appendix we will prove formula (\ref{n2}). One has:
$$
\sum_{i=1}^N\nabla_i^2=\sum_{i=1}^N(\dd_i^2+c\sum_{j=1}^N(\dd_i
\frac{1-\delta_{ij}}{x_i-x_j}{\mathcal{P}}_{ij}+\frac{1-\delta_{ij}
}{x_i-x_j} \dd_j{\mathcal{P}}_{ij})+c^2\sum_{j,k=1}^N\frac{1-
\delta_{ij}}{x_i-x_j}{\mathcal{P}}_{ij} \frac{1-
\delta_{ik}}{x_i-x_k}{\mathcal{P}}_{ik})=
$$
 \be\label{nn2}
=\sum_{i=1}^N(\dd_i^2+c\sum_{j=1}^N\frac{1-\delta_{ij}}{(x_i-x_j)^2
}{\mathcal{P}}_{ij}+c\sum_{j=1}^N\frac{1-\delta_{ij}}{x_i-x_j
}(\dd_i+\dd_j){\mathcal{P}}_ {ij}+c^2
\sum_{j=1}^N\frac{1-\delta_{ij}}{(x_i-x_j)^2}+
 \ee
$$
+c^2\sum_{j,k}^N\frac{(1-\delta_{ij})(1-\delta_{ik})(1-\delta_{kj})}
{(x_i-x_j)(x_i-x_k)}{\mathcal{P}}_{ij}{\mathcal{P}}_{ik}),
$$
then, taking into account symmetry property of permutation operator
(\ref{piks}) one deduces that the third term in (\ref{n2}) vanishes
(as trace of product of symmetric and antisymmetric matrices), while
property (\ref{pikf}) tells that product ${\mathcal{P}}_{ij}{
\mathcal{P}}_{ik}$ remain unchanged under cyclic permutations $i\to
j\to k\to i$ which allows to rewrite the last term in (\ref{n2})
making cyclic transpositions of dummy indices:
$$
\frac13\sum_{i,j,k=1}^N(1-\delta_{ij})(1-\delta_{ik})(1-\delta_
{kj})(\frac1{x_i-x_j}\frac1{x_j-x_k}+\frac1 {x_j-x_k}
\frac1{x_k-x_i}+\frac1{x_k-x_i}\frac1{x_i-x_j}) {\mathcal{P}}_{ij}
{\mathcal{P}}_{ik}=
$$
$$
=\frac13\sum_{i,j,k=1}^N(1-\delta_{ij})(1-\delta_{ik})(1-\delta_{kj}
)\frac{(x_i-x_j)+(x_j-x_k)+(x_k-x_i)}{(x_i-x_j) (x_j-x_k)(x_j-x_k)}
{\mathcal{P}}_{ij}{\mathcal{P}}_{ik}=0.
$$

Now we prove that the Dunkl operator corresponds to flat connection.

Indeed, the commutator $[\nabla_k;\nabla_l]$ consists of four
pieces: the commutator of derivatives (proportional to $c^0$), which
is zero, two pieces, linear by $c$ and one piece is proportional to
$c^2$: $[\nabla_k;\nabla_j]=cI_{kj}^{(1)}+c^2I_{kj}^{(2)}$. One has
due to $j=\neq k$:
$$
I_{kj}^{(1)}=\sum_{i=1}^N((1-\delta_{ij})+\delta_{ij})[\dd_j;\frac{
1-\delta_ {ik}}{x_i-x_k}{\mathcal{P}}_{ik}]-j\leftrightarrow k=(\dd
_j\frac1{x _j-x_k}{\mathcal{P}}_{jk}-\frac1{x_j-x_k}{\mathcal{P}}_
{jk}\dd_j)-j\leftrightarrow k=
$$
$$
=(\frac1{x_j-x_k}(\dd_j-\dd_k){\mathcal{P}}_ {jk}\dd_j-\frac1
{(x_j-x_k)^2}{\mathcal{P}}_{j k})-j\leftrightarrow k=0,
$$
here at first row the commutator at bracket $(1-\delta_{ij})$ i.e.
$i\neq j$ vanishes and in the last row vanishes because the
expression in brackets is symmetric with respect to
$j\leftrightarrow k$.

Consider now $I_{kj}^{(2)}$:
$$
I_{kj}^{(2)}=[\frac{1-\delta_{ik}}{x_i-x_k}{\mathcal{P}}_{ik};
\frac{1-\delta_{lj}}{x_l-x_j}{\mathcal{P}}_{lj}],
$$
and insert there unity:
$$
1=(1-\delta_{il})(1-\delta_{kl})(1-\delta_{ij})+\delta_{il}(1-
\delta_{kl})(1-\delta_{ij})+\delta_{kl}+\delta_{ij}-\delta_{kl}
\delta_{ij},
$$
then first term differs from zero only at $i\neq j\neq k\neq l$,
when factors of commutator commute each to other and it vanishes.
Similarly the last term is also vanishes, because at $i=j$, $l=k$
factors of commutator become equal each to other with opposite sign.
Passing permutation to the right one rewrite remaining three terms
as follows:
$$
\sum_{i=1}^N(1-\delta_{ij})(1-\delta_{ik})(\frac1{x_j-x_k}\frac1
{x_k-x_i}({\mathcal{P}}_{ki}-{\mathcal{P}}_{kj}){\mathcal{P}}_{ij}
+\frac1{x_k-x_i}\frac1{x_i-x_j}({\mathcal{P}}_{ki}-{\mathcal{P}}
_{ij}){\mathcal{P}}_{jk}+
$$
$$
+\frac1{x_j-x_k}\frac1{x_i-x_j}
({\mathcal{P}}_{jk}-{\mathcal{P}}_{ij}){\mathcal{P}}_{ki}),
$$
using identity:
$$
\frac1{x_i-x_k}\left(\frac1{x_i-x_j}+\frac1{x_j-x_k}\right)=\frac1{
x_i-x_j}\frac1{x_j-x_k},
$$
one sees, that all terms canceled due to (\ref{pikf}). So the proof
of statement:
 \be\label{nac}
[\nabla_k;\nabla_l]=0,
 \ee
is finished.
\section{Appendix B}
In  this Appendix we will commute permutation and Dunkl operator.
Consider first the case when all indices different: $j\neq l\neq k$:
$$
{\mathcal{P}}_{jl}\nabla_k={\mathcal{P}}_{jl}[\dd_k-c\sum_{i=1}^N
\frac{1-\delta_{ik}}{x_i-x_k}{\mathcal{P}}_{ik}((1-\delta_{ij})(1-
\delta_{il})+\delta_{ij}+\delta_{il})]=
$$
$$
=[\dd_k-c\sum_{i=1}^N
\frac{1-\delta_{ik}}{x_i-x_k}{\mathcal{P}}_{ik}(1-\delta_{ij})(1-
\delta_{il})]{\mathcal{P}}_{jl}-c{\mathcal{P}}_{jl}(\frac{1-\delta
_{jk}}{x_j-x_k}{\mathcal{P}}_{jk}+\frac{1-\delta_{lk}}{x_l-x_k}
{\mathcal{P}}_{lk})=\nabla_k{\mathcal{P}}_{jl},
$$
here we took into account that $\delta_{ij}\delta_{il}=0$, and in
first term ${\mathcal{P}}_{jl}$ freely moves to right, then we
notice, that extra terms in square bracket just canceled with two
terms standing outside.

Now consider case $l=k$, $j\neq k$:
$$
{\mathcal{P}}_{jk}\nabla_k={\mathcal{P}}_{jk}[\dd_k-c\sum_{i=1}^N
\frac{1-\delta_{ik}}{x_i-x_k}{\mathcal{P}}_{ik}(1-\delta_{ij}+
\delta_{ij})]=[\dd_j-c\sum_{i=1}^N\frac{(1-\delta_{ij})(1-
\delta_{ik})}{x_i-x_k} {\mathcal{P}}_{ij}]{\mathcal{P}}_{jk}-
$$
$$
-c {\mathcal{P}}_{jk}\frac{1-\delta_{jk}}{x_j-x_k}{\mathcal{P}}_
{jk}=\nabla_j{\mathcal{P}}_{jk}.
$$
Here we moved at first step permutation to the right and noticed
that the extra term, coming from $\delta_{ij}$ is just missing term
in sum in square bracket at $i=k$.

\section{Appendix C}
It is seen that the main difficulty with the reduction to normal
form is related to an essential part of the permutation, the
sign-changing operator:
$$
(-1)^{x\dd}f(x)=f(-x).
$$
In order to see it, one can prove at arbitrary complex $q$ the more
general relation:
$$
q^{x\dd_x }f(x)=f(qx).
$$
Indeed, putting $x=e^t$ one has $x\dd_x=\dd_t$ and
$$
q^{\dd_t}f(e^t)=e^{\log q\dd_t}f(e^t)=f(e^{t \log q})=f(qt).
$$
Then, introducing
$$
u=x_j+x_k,\qquad\qquad v=x_k-x_j,
$$
one has
$$
\dd_u=\frac12(\dd_k+\dd_j),\qquad\qquad\dd_v=\frac12(\dd_k-\dd_j)
$$
and
$$
{\mathcal{P}}_{kj}\cdot f(x_k,x_j)=(-1)^{v\dd_v}\cdot f(\frac12(u+v
),\frac12(u-v))=f(\frac12(u+(-v)),\frac12(u-(-v)))=f(x_j,x_k).
$$
The normally ordered expression for ${\mathcal{P}}_{kj}$ is given by
elegant formula (\ref{taylor})
$$
(-1)^{v\dd_v}\cdot f(v)=f(-v)=\sum_{n=0}^\infty\frac{(-2v)^n}{n!}
\dd_v^n\cdot f(v),
$$
which is formally just Taylor expansion of $f(-v)$ around point $v$.
This observation may be replaced by a more lengthy proof, using the
Stirling numbers of the second kind to relate it with (\ref{perm}).

In order to establish the equivalence of relations (\ref{perm}) we
just transform them to normal ordered form, using formula
(\ref{taylor}). Consider:
$$
(-1)^{x_j\dd_j}e^{-x_j\dd_k}e^{x_k\dd_j}e^{-x_j\dd_k}\!=\!\!\!\sum_{n=0}^
\infty\!\!\frac{(-2x_j)^n}{n!}\dd_j^ne^{-x_j\dd_k}e^{x_k\dd_j}e^{-x_j
\dd_k}\!
=\!\!\!\sum_{n=0}^\infty\!\!\frac{(2x_j)^n}{n!}e^{-x_j\dd_k}e^{x_k\dd_j}e^{
-x_j\dd_k}\dd_j^n\!=
$$
$$
=\sum_{n=0}^\infty\frac{(2x_j)^n}{n!}e^{-x_k\dd_j}\sum_{m=0}^\infty
\frac{(x_j)^m}{m!}\dd_k^me^{-x_k\dd_j}\dd_j^n=\sum_{m,n=0}^\infty
\frac{(x_j-x_k)^m}{m!}\frac{(2x_j)^n}{n!}e^{-2x_k\dd_j}(\dd_k-\dd_j)
^m\dd_j^n=
$$
$$
=\!\sum_{m=0}^\infty\frac{(x_k-x_j)^m}{m!}\!\left(\!\sum_{m=0}^
\infty \frac{(2x_j)^n}{n!}\dd_j^n\!\right)\!e^{-2x_k\dd_j}(\dd_k-
\dd_j)^m\!=\! \sum_{m=0}^\infty\frac{(x_k-x_j)^m}{m!}(\dd_k-
\dd_j)^m\!=\!(-1)^{v\dd_v}.
$$
In similar way another expressions for permutation operator via
finite translations and sign-changing operator can be transformed to
normal-ordered form (\ref{taylor}).


\begin{thebibliography}{99}
\bibitem{Ha}Ha, Z.N.C., {\it{Exact dynamical correlation functions
of the Calogero-Sutherland model and one dimensional fractional
statistics in one dimension: View from an exactly solvable model.}}
Nucl. Phys. {\bf{B 435}} (1995), 604–636.

\bibitem{Hal} Haldane, D., {\it{Physics of the ideal fermion gas:
Spinons and quantum symmetries of the integrable Haldane-Shastry
spin chain.}} In: A. Okiji, N. Kamakani (eds.), Correlation effects
in low-dimensional electron systems. {\it{Springer}}, 1995, pp.
3–20.

\bibitem{cal}F. Calogero, {\it{Solution of a three-body problem in
one dimension}}, J. Math. Phys. 10 (1969), 2191-2196\\ F. Calogero,
{\it{Solution of the one-dimensional N-body problem with quadratic
and/or inversely quadratic pair potentials}}, J. Math. Phys. 12
(1971), 419-436\\ F. Calogero, {\it{Exactly solvable one dimensional
many-body problems}}, Lett. Nuovo Cimento {\bf{13}} (1975), 411-416

\bibitem{p}A.~P.~Polychronakos, {\it{Exchange operator formalism for
integrable systems of particles.}}Phys.Rev.Lett. {\bf{69}}
(1992),703–705.\\ For rewiev see also:  A.P.~Polychronakos, J. Phys.
A {\bf 39} (2006) 12793.

\bibitem{He} G.~J.~Heckman, {\it{A remark on the Dunkl
differential-difference operators.}} In: Barker, W., Sally, P.
(eds.) Harmonic analysis on reductive groups. Progress in Math. 101,
Birkh\"auser, 1991. pp. 181 – 191.

\bibitem{Pe}A.~M.~Perelomov, {\it{Algebraical,  approach to the
solution of a one-dimensional model of N interacting particles.}}
Teor. Mat. Fiz. {\bf{6}} (1971), 364–391.

\bibitem{Mo} J.~Moser, {\it{Three integrable Hamiltonian systems
connected with isospectral deformations,}} Adv. in Math. {\bf{16}}
(1975), 197–220.

\bibitem{dunkl} Ch. F. Dunkl, (1989), {\it{Differential-difference
operators associated to reflection groups}}, Transactions of the
American Mathematical Society 311 (1): 167–183

\bibitem{dun} C.~F.~Dunkl, {\it{Reflection groups and orthogonal
polynomials on the sphere.}} Math. Z. {\bf{197}} (1988), 33–60.\\
C.~F.~Dunkl, {\it{Operators commuting with Coxeter group actions on
polynomials.}} In: Stanton, D. (ed.), Invariant Theory and Tableaux,
Springer, 1990, pp. 107–117.\\
C.~F.~Dunkl, {\it{Integral kernels with reflection group invariance.
}} Canad. J. Math. 43 (1991), 1213–1227.\\
C.~F.~Dunkl, {\it{Hankel transforms associated to finite reflection
groups.}} In: Proc. of the special session on hypergeometric
functions on domains of positivity, Jack polynomials and
applications. Proceedings, Tampa 1991, Contemp. Math. 138 (1992),
pp. 123–138.

\bibitem{hec}Heckman, {\it{Dunkl operators. S´eminaire Bourbaki}}
828, 1996–97; Ast´erisque 245 (1997), 223–246

\bibitem{ef} G.~V.~Efimov, {\it{The non-local interactions}}, Nauka,
Moskow 1977 (in russian).

\bibitem{kbi} Korepin, V. E.; Bogoliubov, N. M.; Izergin, A. G.
(1993), {\it{Quantum inverse scattering method and correlation
functions, Cambridge Monographs on Mathematical Physics}}, Cambridge
University Press, ISBN 978-0-521-37320-3, MR 1245942

\bibitem{skl}E. K. Sklyanin, {\it{Dynamical r–matrices for elliptic
Calogero–Moser model}} Algebra and Analysis (1994) 227-237,

\bibitem{tsu} N.~Yonezawa, I.~Tsutsui, {\it{Inequivalent
Quantizations of the N = 3 Calogero model with Scale and
Mirror-$S_3$ Symmetry}}, J.Math.Phys. {\bf{47}} (2006) 012104

\bibitem{tsu1}L.~Feher, I.~Tsutsui, T.~Fulop, {\it{Inequivalent
quantizations of the three-particle Calogero model constructed by
separation of variables}}, Nucl.Phys. {\bf{B715}} (2005) 713-757.
\end{thebibliography}
\end{document}